\documentclass{Interspeech2024}
\usepackage{algorithm}
\usepackage{algpseudocode}
\usepackage{array}
\usepackage{amsmath}
\usepackage{comment}
\usepackage{multirow}
\usepackage{graphicx}
\usepackage{subfig}
\newcommand\numberthis{\addtocounter{equation}{1}\tag{\theequation}}

\interspeechcameraready

\title{CTC-aligned Audio-Text Embedding for Streaming Open-vocabulary Keyword Spotting}

\name{Sichen}{Jin}
\name{Youngmoon}{Jung}
\name{Seungjin}{Lee}
\name{Jaeyoung}{Roh}
\name{Changwoo}{Han}
\name{Hoonyoung}{Cho}

\address{
  Samsung Research, South Korea} 
\email{\{sc.ehkim.jin, youngm.jung, sjsr.lee, jyo.roh, cw1105.han, h.y.cho\}@samsung.com}


\begin{document}

\maketitle

\begin{abstract}
This paper introduces a novel approach for streaming open-vocabulary keyword spotting (KWS) with text-based keyword enrollment. 
For every input frame, the proposed method finds the optimal alignment ending at the frame using connectionist temporal classification (CTC) and aggregates the frame-level acoustic embedding (AE) to obtain higher-level (i.e., character, word, or phrase) AE that aligns with the text embedding (TE) of the target keyword text. After that, we calculate the similarity of the aggregated AE and the TE.
To the best of our knowledge, this is the first attempt to dynamically align the audio and the keyword text on-the-fly to attain the joint audio-text embedding for KWS. 
Despite operating in a streaming fashion, our approach achieves competitive performance on the LibriPhrase dataset compared to the non-streaming methods with a mere 155K model parameters and a decoding algorithm with time complexity $O(U)$, where $U$ is the length of the target keyword at inference time.

\end{abstract}

\section{Introduction}


Keyword Spotting (KWS) plays an important role in voice assistants by detecting wake-up words and enabling hands-free activation. 
Over the years, KWS has developed from fixed vocabulary \cite{traditional,tara_small,google_small}, where the models are trained with large amount of audio data of the same keywords, to open-vocabulary systems where users are allowed to use customized keywords by going through the enrollment process without re-training the models. Open-vocabulary KWS can be challenging since it requires the system to be flexible and robust to unseen keywords.

There are two possible ways of enrollment for customized keywords. One is named Query-by-Example (QbyE) \cite{qbye_attention,qbye_google,donut}, where several example keyword utterances are provided. The similarities between the input queries and the enrolled examples are detected by aligning the frame-level acoustic embedding (AE) using Dynamic Time warping (DTW) \cite{donut} or attention \cite{qbye_attention}. These methods work well in controlled environments. However, their reliance solely on vocal features makes them susceptible to reduced performance under varying conditions and across diverse user voices. 
On the other hand, the enrollment process of speaking the same keyword several times can feel inconvenient and unnatural.

Another way of enrollment can be as simple as inserting keywords in text form. This intuitive method can significantly enhance the user interface for its simplicity.
There has been several approaches to detect the spoken keywords by mapping both text and speech to a shared latent encoding space. To handle the difference in the lengths of the keywords and their spoken form, attention \cite{libriphrase, phonmatchnet} and alignment methods based on dynamic programming \cite{apple1, apple2} were explored to match the corresponding text and audio frames sharing the same syntax. While these methods are showing promising results, they often need computations on the global context, making them hard to run in a streaming manner. 

As the front-line user interface of voice assistants, the computations for KWS need to be very small and use least possible information from the past in order for them to be active all the time. Convolutional neural networks (CNN) \cite{cnn,mobilenet} are often used in small-footprint KWS systems \cite{tara_small,google_small} to yield better performance with smaller models. Connectionist temporal classification (CTC) \cite{ctc,ctc_chinese} is employed to get phoneme hypotheses for each audio frame for streaming KWS. 
Building upon these works, it occurred to us that integrating embedding-based methods could elevate small-footprint online KWS to a new level.

In light of the requirements and the difficulties, we propose a novel structure combining the advantages of CTC and embedding approaches: CTC for instant information retrieval from each audio frame and embedding for comparing the global information of the entire keyword. With CTC-aligned Audio-Text (CTCAT) keyword detector, we provide an end-to-end training strategy to learn the CTC alignment and the joint embedding space simultaneously, together with an inference algorithm with time complexity $O(U)$ where $U$ is the keyword length.
Experimental results on the LibriPhrase dataset show that the proposed KWS system achieves performance comparable to that of its non-streaming counterparts, despite having a significantly smaller model size (i.e., an acoustic encoder with just 155K parameters) and faster inference times.

\begin{figure*}[t]
  \centering
  \vspace{-0.5cm}
  \includegraphics[width=\textwidth]{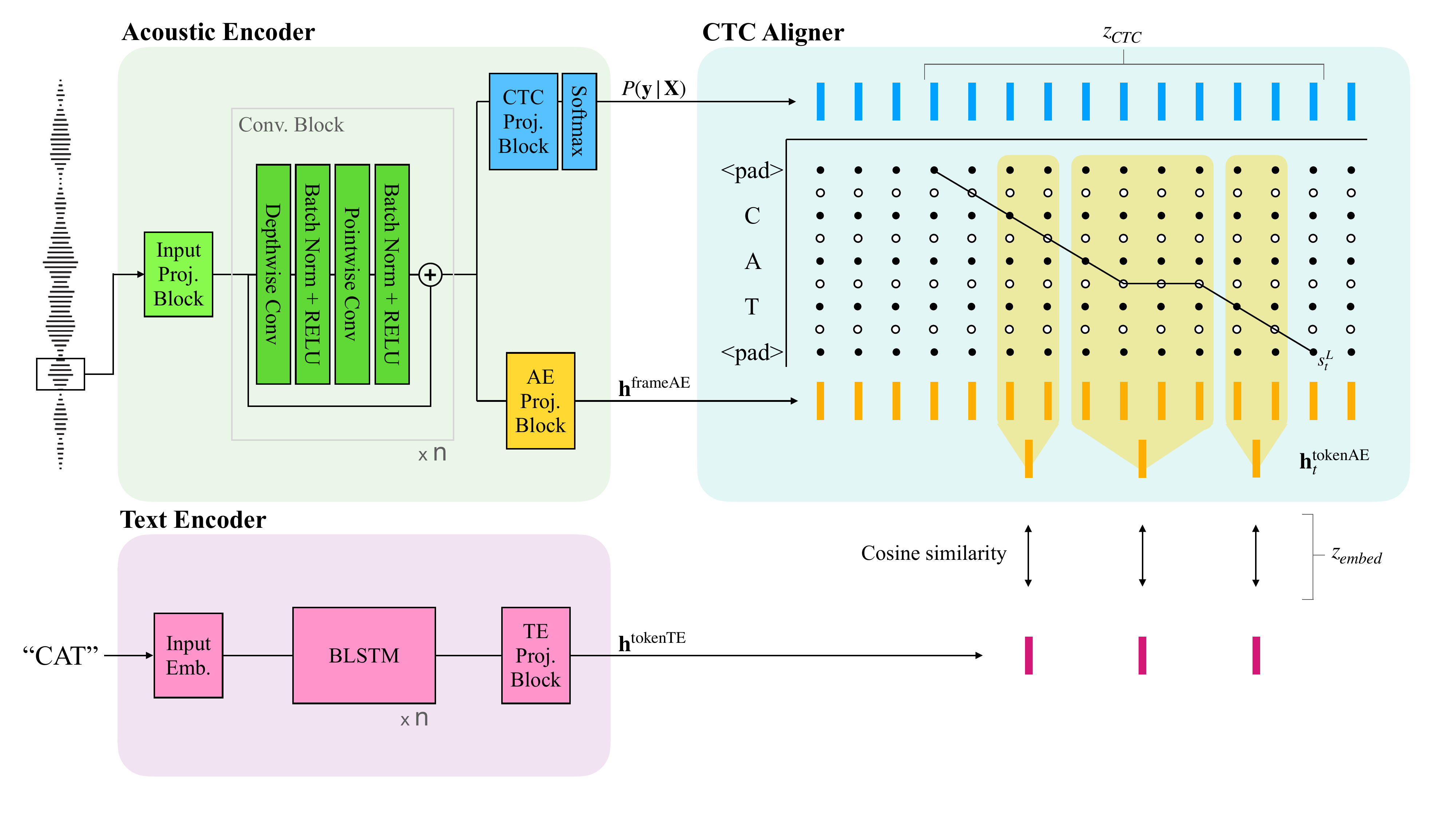}
  \vspace{-1.0cm}
  \caption{The depiction of the entire structure of the CTC-aligned Audio-Text (CTCAT) keyword detector. The acoustic encoder and the text encoder encodes the input audio sequence and the target keyword tokens into acoustic embedding (AE) and text embedding (TE), and the CTC aligner aligns the embedding vectors that share the same tokens.}
  \vspace{-0.5cm}
  \label{fig:structure}
\end{figure*}

\section{Related Work}
In this work, we attempt to merge the strengths of streaming and non-streaming KWS techniques, introducing an innovative approach that leverages both the frame-wise information using CTC and the global context using multi-view loss.
We explain these two methodologies in the following section.

\subsection{CTC}
CTC \cite{ctc} is among the earliest attempts in ASR to automatically learn the alignment between the speech frames and the transcript without frame-wise labels. To deal with the length difference between the transcript and the spoken utterance, a new blank token is added to the original vocabulary, making \(V'=V \cup \{Blank\}\). At each time step \(t\), the output of the neural network \(y_k^t\) is interpreted as the probability observing label \(k\) at time \(t\). 
$V'^T$ is the set of length $T$ sequences over the $V'$ and $\pi$ is defined as an element of $V'^T$.
The distribution of one possible path $\pi$ of time \(T\) can be interpreted as the product of the probabilities of each token $\pi_t$ on the way.

\begin{equation}
p(\pi | \mathbf{X}) = \prod^T_{t=1} y^t_{\pi_t}, \forall \pi \in V'^T
\end{equation}

In order to compute the probability of the target labels, a many-to-one map $\mathcal{B}$ is defined where all blank tokens and repeated labels from the paths are removed. Finally, the probability of the target label \(\mathbf{l}\) is defined as the sum of the probabilities of all the paths corresponding to it, as follows: 

\begin{equation}
p(\mathbf{l}|\mathbf{X}) = \sum_{\pi \in \mathcal{B}^{-1}(\mathbf{l})} p(\pi|\mathbf{X}).
\end{equation}

The forward-backward algorithm is used to efficiently calculate the probability \(p(\mathbf{l} | \mathbf{X})\) and the loss function is defined as the log probability of the target label: 

\begin{equation}
\label{eq_CTC}
L_{\textrm{CTC}} = -ln(p(\mathbf{l} | \mathbf{X})).
\end{equation}

\subsection{Multi-view Loss} 
Jung \textit{et al}. \cite{Jung-INTERSPEECH} extended the multi-view learning method \cite{He-ICLR} using the asymmetric proxy loss (AsyP) loss by setting the text embedding (TE) as proxies. We refer to the AsyP loss as `multi-view loss' throughout this paper, and we explain the loss using the same notations as in \cite{Jung-INTERSPEECH}.
Let a mini-batch consist of a set of $N$ data tuples, which is denoted as $\{(\bm{x}_i, \bm{t}_i, c_i)|i=1,2,\cdots,N\}$, where $\bm{x}_i$ is the AE of the $i$-th speech segment, $\bm{t}_i$ is the TE of the corresponding text, and $c_i$ is the word label.
The multi-view loss is defined as the sum of the anchor--positive (AP) and anchor--negative (AN) terms, as follows:
\begin{align*}
    \mathcal{L}_\textrm{multi-view}=\frac{1}{N} \sum_{i=1}^N \bigg (\frac{1}{\alpha} \underset{j \in \mathcal{P}_i}{\textrm{ELSE}} \, \alpha(\lambda - S(\bm{t}_i,\bm{a}_j)) \\ + \underset{k \in \mathcal{N}_i}{\textrm{MSP}} \, \beta(S(\bm{a}_i,\bm{t}_k) - \lambda) \bigg), \numberthis \label{eq_asyp}
\end{align*}
where $\mathcal{P}_i = \{j|y_j=y_i\}$, $\mathcal{N}_i = \{k|y_k \neq y_i\}$, and $S(\cdot,\cdot)$ are the set of positive indices, the set of negative indices, and the cosine similarity, respectively. $\textrm{ELSE}$ and $\textrm{MSP}$ are the Extended-LogSumExp \cite{Wang-CVPR} and Mean-Softplus \cite{Yi-ICPR} functions, respectively.
The AP term draws anchor $\bm{t}$ and positives $\bm{a}$ closer, while the AN term pushes the anchor $\bm{a}$ and negatives $\bm{t}$ apart. 
$\alpha$, $\beta$, and $\lambda$ are hyper-parameters determining the boundaries in the embedding space or the severity of penalty for violations.

\section{Proposed Method}

\subsection{Architecture}
The overall system consists of three main parts: a text encoder, an acoustic encoder and a CTC aligner as shown in Fig. \ref{fig:structure}.

The text encoder encodes the keyword text to latent vectors.
First, the text is tokenized into smaller units, such as characters, as the CTC training requires. Then, the tokenized keyword $\mathbf{y} = \{y_u |1 \leq u \leq U\}$ are fed into two bi-directional LSTM layers to get the token-level TE \(\mathbf{h}^{\mathrm{tokenTE}}_{y_u}\), where $u$ is the index of non-blank tokens and $U$ is the total number of non-blank tokens. 

The acoustic encoder operates in streaming mode to encode the input audio stream \(\mathbf{X}=\{\mathbf{x}_t | 1 \leq t \leq T \} \) to latent vectors.
After that, the latent vectors are fed into two projection blocks: CTC and AE projection blocks, resulting in token distributions \(P(\mathbf{y} | \mathbf{x}_t)\) and frame-level AE \(\mathbf{h}^{\mathrm{frameAE}}_t\) for each frame \(t\), respectively.
In this work we stack mobilenet \cite{mobilenet} blocks in order to keep the size small. 

\begin{figure}[t!]
  \centering
  \includegraphics[width=\linewidth]{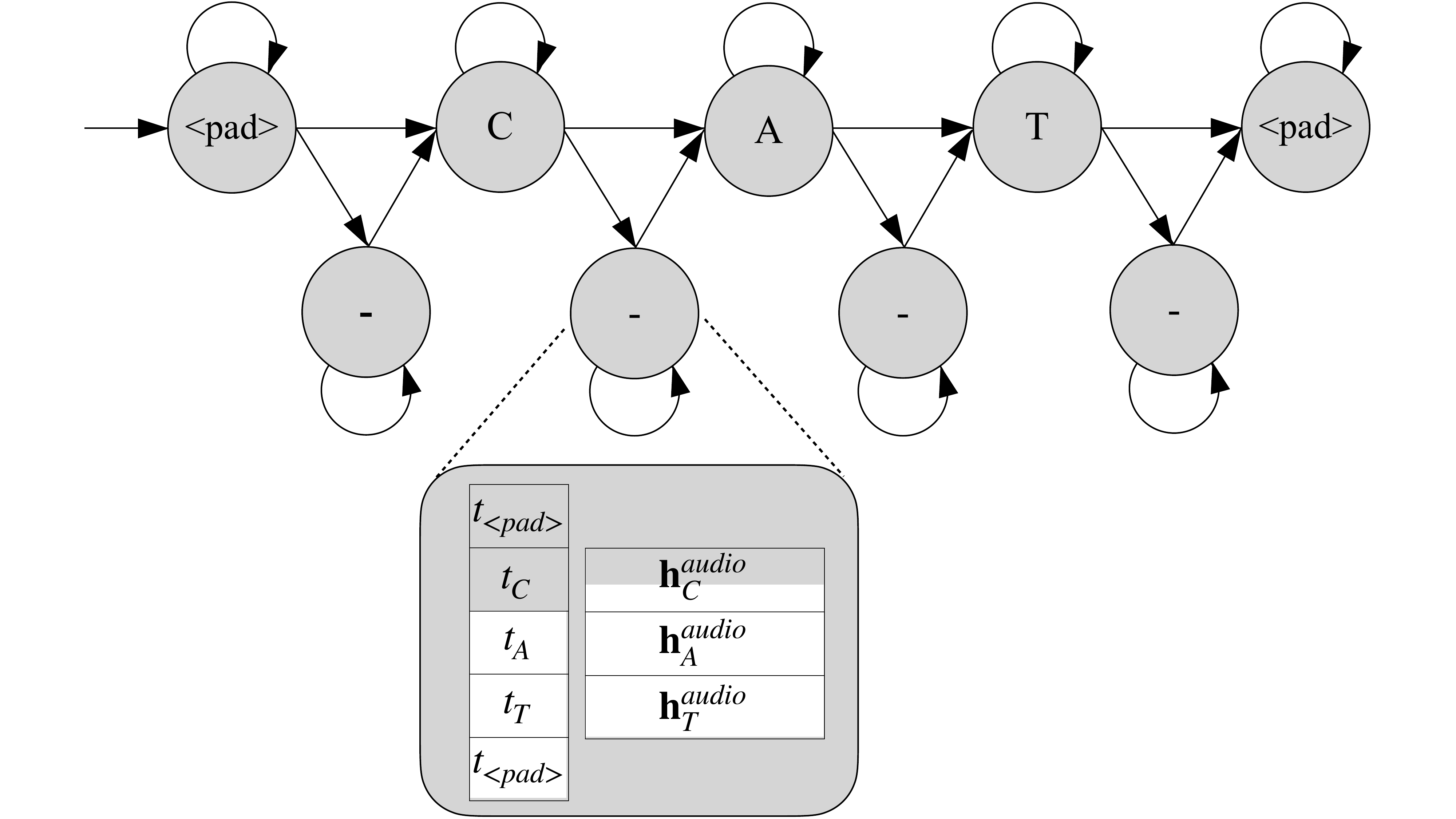}
  \caption{The decoding graph made for the target keyword ``cat". In each state, transition timings and the accumulated AE are stored. The portions colored white in the state means the information is not (half) filled.}
  \label{fig:decoding_graph}
  \vspace{-0.5cm}
\end{figure}

The structure of the CTC aligner is shown in Fig. \ref{fig:decoding_graph}. The states of the decoding graph \(\{s_l | 1 \leq l \leq L\}\) each corresponds with the keyword tokens $\mathbf{y}$ adding the blank tokens in between \(\{y_1, -, y_2, -, ..., -, y_U\}\), where $l$ is the state index and \(L = 2U-1\). We will refer to the token corresponding to a state $l$ as $st_l$. Then, we can say that $st_l = y_{2u-1}$ and $y_u = st_{\lceil \frac{l}{2} \rceil} $ for each non-blank token $u$. The notations will interchange throughout the paper. Each state stores the accumulated CTC score \(z^{l,t}_{\textrm{CTC}}\), the transition timing information when the previous non-blank states on the path were first visited \(\mathbf{t}^{l,t}_{y_{1:U}}\), and the accumulated AE for the non-blank tokens \(\mathbf{h}^{l,t}_{y_{1:U}}\). Note that the graph both starts and ends with non-blank tokens unlike the usual forward-backward algorithm since we are interested only in the portion of the audio where the keyword is pronounced, rather than aligning the entire input audio sequence with the given keyword.

In practice, we add padding tokens to the beginning and the end of the keyword tokens when training the CTC aligner. 
We found this helps the model converge with lower CTC losses and the extra padding token at the end ensures each non-blank token accumulates its corresponding frame-wise embeddings without any missing.
The blank tokens following a non-blank token are considered as the continuation of the pronunciation of the non-blank token and thus the frame-level AE of the blank tokens is added to the token-level AE of the leading non-blank token.
It is worth mentioning that the padding tokens are only used in algorithms related to CTC alignments, whereas only the non-blank keyword tokens are considered for the comparison of the matching AE and TE.


\subsection{CTC Aligner}
\label{section:ctc_aligning}


At every time step \(t\), the transition probabilities of the decoding graph are updated according to the output distribution over the vocabulary \(V^* = V \cup \{Padding, Blank\}\) calculated from the CTC projection block. 
Then we can use the Viterbi algorithm \cite{viterbi} to get the best forced-aligned paths for the target keyword that ends at the current frame. 
As the first state \(s_1\) is always set independent of other states, its accumulated score and transition timing information are defined as follows: 
\begin{equation}
\begin{split}
z^{1,t}_{\textrm{CTC}} &= p(y_1|\mathbf{x}_t), \\
t^{1,t}_{y_1} &= t.
\end{split}
\end{equation}

Since the probabilities of the incoming transitions of a state are all defined as the probability of the corresponding token of the state at the time, we can simply compare the scores of the upstream states and decide the source state. 
\begin{equation}
I^{l,t} = 
    \begin{cases} 
        \mathop{arg\,max}\limits_{i \in \{l, l-1\}} z^{i,t-1}_{\textrm{CTC}} & \text{if } l \text{ is even (blank)} \\
        \mathop{arg\,max}\limits_{i \in \{l, l-1, l-2\}} z^{i,t-1}_{\textrm{CTC}} & \text{if } l \text{ is odd (non-blank)}
    \end{cases}
\end{equation}

The transition timings and the accumulated AE are inherited from the source state. For states corresponding to the non-blank tokens, the transition timing for the corresponding token $t^{l,t}_{st_l}$ is updated as the current time \(t\), and the frame-level AE $\mathbf{h}^{\mathrm{frameAE}}_t$ is added to the accumulated AE for the corresponding token $\mathbf{h}^{l,t}_{st_l}$. For states corresponding to blank tokens, the frame-level AE is added to the last non-blank token $\mathbf{h}^{l,t}_{st_{l-1}}$. The score of the state is updated as follows:
\begin{equation}
z^{l,t}_{\textrm{CTC}} = z^{I^{l,t},t-1}_{\textrm{CTC}} + logP(st_l|\mathbf{x}_t).
\end{equation}

The proposed aligner uses dynamic programming and does not require storing and revisiting of the computation histories. For decoding only, the time complexity required for one time step is $O(U)$ and the space complexity is $O(U^2)$ where $U$ is the length of the keyword.
Since the length of the keyword is much smaller than the length of audio input sequences, the algorithm is much more efficient than the non-streaming algorithms.

\subsection{Training Strategy}
Without the need of a pre-trained aligner, 
we train the model end-to-end from scratch to ensure the acoustic encoder learns both the CTC alignment and the AE at the same time. During training however, the entire audio input sequence of finite length is provided at once unlike the inference time. We skip storing the AE for the storation itself will consume $O(TU^2)$ memory space, and only save the CTC scores and the transition timings of the final state for every time step in the training sample. After the alignment is finished, we choose the aligned keyword path at $t_{optimal}$ with the highest CTC score among all the frames. The token-level AE for each non-blank token $y_u$ is obtained by pooling the frame-level AE between the transition timings stored in the state $t_{y_u} = t^{L,t_{optimal}}_{y_u}$:

\begin{equation}
\mathbf{h}^{\mathrm{tokenAE}}_{y_u} = \dfrac{\sum_{f=t_{y_u}}^{t_{y_{u+1}}-1} \mathbf{h}^{\mathrm{frameAE}}_f}{t_{y_{u+1}} - t_{y_u}}.
\end{equation}
The objective function is defined as a combination of the $L_{\textrm{CTC}}$ in Eq. (\ref{eq_CTC}) and the $\mathcal{L}_\textrm{multi-view}$ in Eq. (\ref{eq_asyp}), which measures the difference between the aligned AE and the TE:
\begin{equation}
L = L_{\textrm{CTC}} + L_{\textrm{multi-view}}(\mathbf{h}^{tokenAE}_{y_{1:U}}, \mathbf{h}^{tokenTE}_{y_{1:U}}).
\end{equation}

In this work, we use character as the base tokens. For the Libriphrase \cite{libriphrase} dataset, we also try to combine the embedding vectors of multiple characters to form higher level embedding: word and phrase levels. For word level, the frame-level AE and the token-level TE are further accumulated until the space character is encountered. Whereas for phrase level, the embedding vectors of the entire keyword are all pooled to make one keyword embedding vector.


\subsection{Inference}

Notably, there is no need to re-visit the results from previous time steps, for they are preserved in the final state via dynamic programming. Every time step, all we need is to retrieve the AE from last state of the updated graph and calculate the cosine similarities between the AE and TE of the target keyword:

\begin{equation}
\begin{split}
z_{\textrm{CTC}} &= z^{l,t}_{\textrm{CTC}}, \\
z_{\textrm{embed}} &= \dfrac{1}{U} \sum_{u=1}^{U} cos(\mathbf{h}^{tokenAE}_{y_u}, \mathbf{h}^{tokenTE}_{y_u}). \\
\end{split}
\end{equation}

The final score can then be derived by linearly integrating the two scores:
\begin{equation}
\label{eq:final}
z = z_{\textrm{CTC}} + \lambda z_{\textrm{embed}},
\end{equation}
where $\lambda$ is a hyper-parameter that needs to be tuned on a development set.

\section{Experiment Result}

We trained our CTCAT keyword detector with an acoustic encoder with only 155K parameters. The detailed structure can be found in Fig. \ref{fig:structure}, where 12 Convolutional blocks were stacked whose kernel size for the depthwise convolution and the feature map number for the pointwise convolution being 12 and 96. 
All the projection blocks consisted of one dense layer followed by a batch normalization layer \cite{batchnorm}. For the text encoder, the input tokens were turned into 256-dimension vectors by a trainable look-up table and then fed into two BLSTM layers with hidden size of 256. We used 28 tokens (English alphabets, space, apostrophe) to train the text encoder, and added two more tokens for padding and blank to train the CTC aligner. 80-channel filterbanks were extracted from the input audio stream with a window of 25ms and a stride of 10ms. 
For the multi-view loss in Eq. (\ref{eq_asyp}), we set $\alpha = 2$, $\beta = 50$, and $\lambda = 0.1$.
Our mini-batches contained 1024 keywords, each of which was comprised with two examples as the positive pair, and the examples for all other keywords acted as the negative example. Adam optimizer \cite{adam} and cosine decay \cite{cosine_decay} are used until 100 epochs was reached for the Libriphrase training set. It took us two days to train the models on two A100 GPUs \cite{a100}.

\begin{table}[t!]
  \vspace{-0.2cm}
  \caption{EER (\%) and AUC (\%) on the Libriphrase evaluation set. \textbf{CTC} is trained only with the CTC loss. The aligned embedding vectors are trained in three different levels: character, word, and phrase level. `\#Params' denotes the (estimated) number of parameters of models used for inference.}
  \vspace{-0.2cm}
  \label{tab:full}
  \centering
  
\begin{tabular}{lllllll}
\hline
\multirow{2}{*}{Method}  & \multirow{2}{*}{\#Params}    & \multicolumn{2}{l}{EER (\%)}    & \multicolumn{2}{l}{AUC (\%)}     \\
                 &           & $\mathbf{LP_E}$       & $\mathbf{LP_H}$        & $\mathbf{LP_E}$        & $\mathbf{LP_H}$        \\ 
\hline
Attention \cite{libriphrase} & $\sim$420K & 8.24          & 32.90          & 96.70          & 73.58          \\
DSP \cite{apple1}            & 3.7M  & 7.36          & \textbf{23.36} & 97.83          & \textbf{84.21} \\ 
\hline
CTC               & 147K       & 9.11         & 32.37          & 96.76          & 73.95          \\
+character        & 155K       & 8.65         & 32.76         & 96.99          & 73.53          \\
+word             & 155K       &   7.05       &    31.62       &   97.97       &   75.12     \\
+phrase           & 155K       & \textbf{6.06} & 29.63         & \textbf{98.32} & 77.10 \\      
\hline
\end{tabular}
\vspace{-0.5cm}
\end{table}

The Libriphrase \cite{libriphrase} dataset was chosen for evaluation for a fair comparison with other methods. Our system was trained from scratch with the training set from Libriphrase which was generated from \textit{train-clean-100} and \textit{train-clean-360} with phrases with 1 to 4 words. 
To make our model more robust, we convolved the clean speech signals with synthetic room impulse responses (RIRs) from the OpenSLR dataset \cite{Ko-ICASSP} and added the MUSAN noise dataset \cite{Snyder-arxiv} at randomly selected signal-to-noise ratios (SNRs) between -3 and 25 dB. 
The evaluation set was generated from \textit{train-others-500} and the negative examples are divided into two sets, easy ($\mathbf{LP_E}$) and hard ($\mathbf{LP_H}$), based on Levenstein distances. \cite{ld} We used the same evaluation dataset which contains 4391, 2605, 467, and 56 episodes for anchor phrases of each length. Each episode contains 3 positive and 3 negative pairs. The text of the anchors were used for text enrollment and the audio of the comparison examples are used for the evaluation of the keyword detection task.
The $\lambda$ in Eq. \ref{eq:final} was decided from 100k example that we randomly held out from the training set. The final results were shown with $\lambda=6$.

We compared our system with the non-streaming open-vocabulary keyword spotting systems which used the cross-attention \cite{libriphrase} and a dynamic programming based algorithm called Dynamic Sequence Partitioning (DSP) \cite{apple1} to align the input audio sequence and the target text. Using the combined scores of CTC and embedding, our system outperformed the non-streaming solutions on $\mathbf{LP_E}$ set, and showed competitive results on $\mathbf{LP_H}$ set. This is a very encouraging result considering the extremely small acoustic encoder with 155K parameters and 6.91M FLOPs per time step and the highly efficient inference algorithm with time complexity $O(U)$. The detailed result is shown in Table \ref{tab:full}.

Four models: trained with only CTC loss, adding joint embedding learnings for token (character), word and phrase levels are all trained from scratch individually. From the experiment results in Table \ref{tab:full}, we can confirm that using the embedding score on top of the CTC score enhances the performance by utilizing the global context over the entire keyword. Moreover,
a slight performance drop on $\mathbf{LP_H}$ was noticed when character-level embedding score was adopted whereas using higher level abstraction resulted in huge performance gain. This indicates that considering more global context help segregate the embedding vectors in the latent space better. Table \ref{tab:length} shows the average number of frame-level AE accumulated to form a single AE in Libriphrase evaluation set.

\vspace{-0.2cm}
\begin{table}[th]
\caption{The average number of frame-level acoustic embedding (AE) accumulated to form a single higher level AE.}
\vspace{-0.2cm}
\label{tab:length}
\centering

\begin{tabular}{llll}
\hline
               & character &  word   &  phrase  \\
\hline
average length &   5.6     &  31.7   &   49.2    \\
\hline
\end{tabular}
\vspace{-0.2cm}
\end{table}

\begin{figure}[t!]%
    \vspace{-0.5cm}
    \centering
    \subfloat[\centering Audio-text embedding]
    {{\includegraphics[width=.5\linewidth,trim={2cm 0 2.5cm 1.5cm}]
    {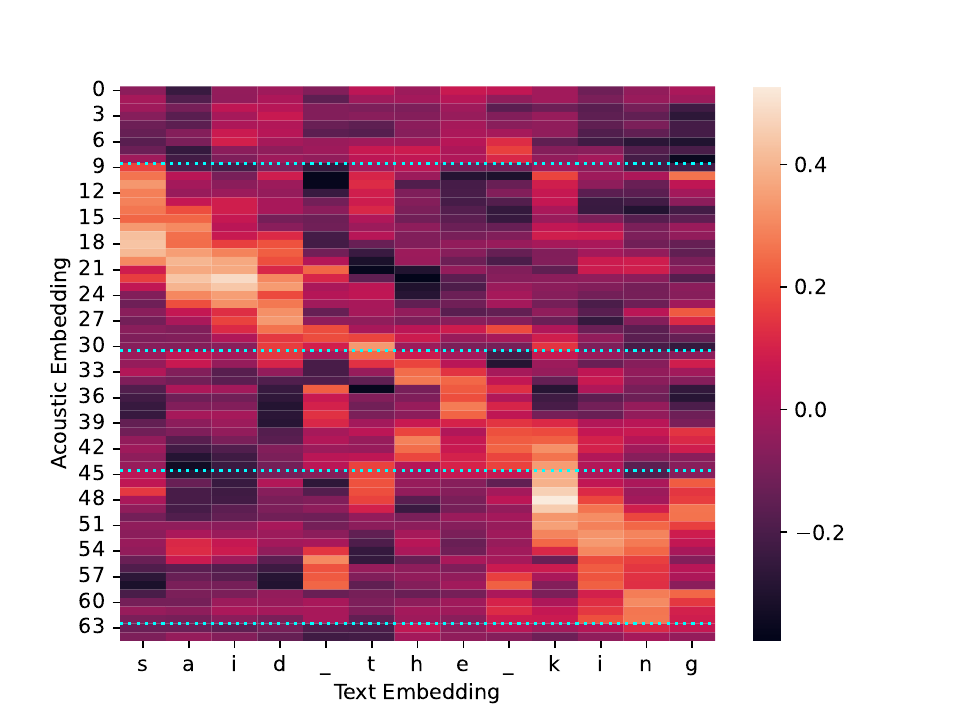} }}%
    \subfloat[\centering Acoustic embedding]
    {{\includegraphics[width=.5\linewidth,trim={2cm 0 2.5cm 1.5cm}]
    {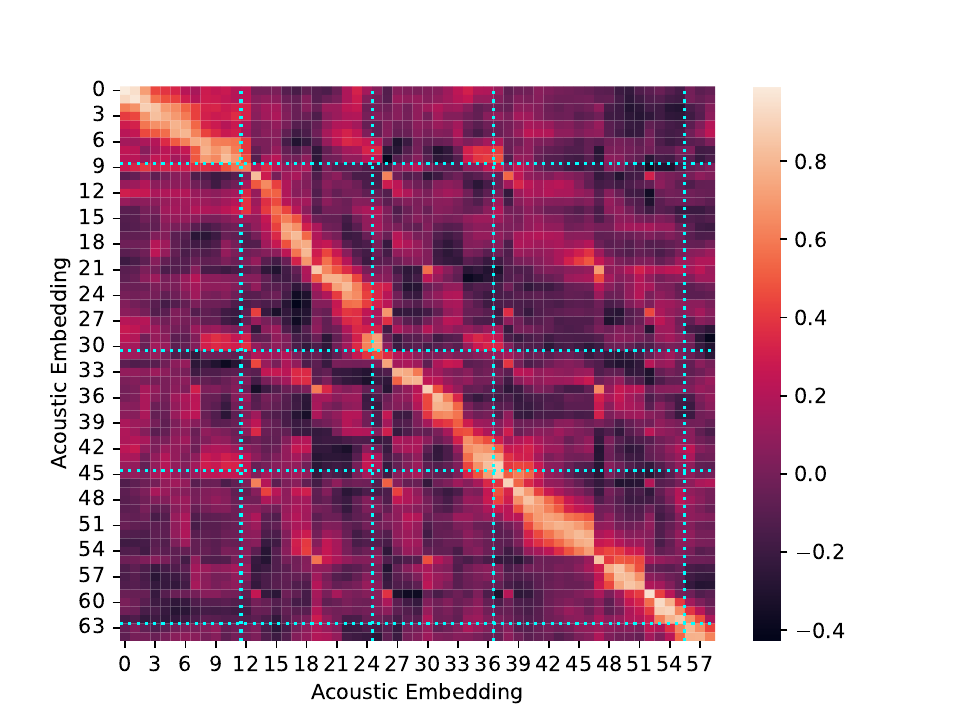} }}%

    \vspace{-0.2cm}
    
    \caption{The correlation map between the audio-text embedding of the positive examples for the target keyword ``said the king''. The blue lines denote the aligned result from CTC for the audio sections of $ \langle silence \rangle , ``said",``the", ``king", \langle silence \rangle$}%
    
    \vspace{-0.5cm}
    \label{fig:visual}%
\end{figure}

We also visualized the correlations between the embedding vectors and the CTC alignments at the same time to confirm the benefit of our joint training strategy in Fig \ref{fig:visual}. We chose the model trained with the word-level embedding to see both the correlations between the frame-level AE within the same word and between different words. Higher correlations were found within the aligned paths which indicates that the CTC score and the embedding score can work coherently without interfering with each other.

\section{Conclusion}
We have introduced a novel CTC-aligned Audio-Text (CTCAT) keyword detector to obtain the joint audio-text embedding in streaming KWS.
We achieved dynamic aligning between the target keyword text and the input audio sequence using CTC, and used the acoustic embedding along the paths to obtain the embedding similarities. Experiments on the LibriPhrase dataset showed that our method shows competitive performance compared to non-streaming methods with much smaller model size and time consumption.

\bibliographystyle{IEEEtran}

\end{document}